# Magnetic, magneto-thermal and magneto-transport properties in SmMn$_2$Si$_{2-x}$Ge$_x$ compounds

Pramod Kumar, K. G. Suresh*

*Department of Physics, I. I. T .Bombay, Mumbai 400076, India*

A. K. Nigam and S. K. Malik

*Tata Institute of Fundamental Research, HomiBhabha Road, Mumbai 400005, India*

Abstract

The effect of Ge substitution for Si in SmMMn$_2$Si$_{2-x}$Ge$_x$ compounds has been studied. The Sm ordering temperature is found to be much larger in the compound with x=2, as compared to the compounds with x=0 and 1. The increase in the intra layer Mn-Mn distance is found to be responsible for this increase. Among these three compounds, SmMn$_2$Ge$_2$ is found to show re-entrant ferromagnetism at low temperatures. The magnetic contribution to the heat capacity has been found in all the three compounds. The splitting of the ground state multiplet has been estimated by fitting the magnetic part of the heat capacity data using the Schottky formula. The isothermal magnetic entropy change is found to remain the same for x=0 and 1, but decrease in the compound with x=2, though the nature of magnetic transition changes from second order to first order, as x is increased from 0 to 2. The electrical resistivity increases with Ge concentration. The excess resistivity in the antiferromagnetic region has been calculated.

**PACS :** 75.30Sg, 75.50Ee,75.50Gg

*Corresponding author (email:suresh@phy.iitb.ac.in)

# 1. INTRODUCTION

Rare earth-based intermetallic compounds draw considerable attention recently owing to their potential for various applications. One of the most important applications being considered in the near future is magnetic refrigeration[1-5]. The underlying physical property of a magnetic refrigerant is the magnetocaloric effect (MCE). In response to a magnetic field, MCE manifests as isothermal magnetic entropy change ($\Delta S_M$) or adiabatic temperature change ($\Delta T_{ad}$), depending on the thermodynamic conditions. Recently, 'near room-temperature magnetic refrigeration technology' has received global attention by virtue of its environment-friendly nature, better adaptability and higher energy efficiency. The discovery of the giant MCE in some materials has given an additional impetus for the study of novel and potential materials with similar properties. Giant MCE has been found in systems such as $MnAs$[6] and $LaFe_{11.4}Si_{1.6}$[7,8], which show first-order ferromagnetic to paramagnetic transition. In addition to the potential use of magnetocaloric materials in magnetic refrigerators, many of these materials also display interesting magnetic and related properties.

One important class of materials being probed recently is $RMn_2X_2$ (X=Si, Ge) series. These compounds in general crystallize in the tetragonal $ThCr_2Si_2$-type structure. They possess layered structure with the sequence Mn-X-R-X-Mn, with the layers perpendicular to the c-axis. An interesting feature of the $RM_2X_2$ compounds with different 3d transition metals (M=Mn, Fe, Co, Ni) is that only for M=Mn, the transition metal sublattice has a nonzero magnetic moment. Some of these compounds exhibit phenomena such as multiple magnetic transitions, first order transitions (FOT), re-entrant ferromagnetism etc[9-15]. Another important feature of these compounds is the strong dependence of structural parameters on the magnetic properties. It has been reported that in most of these compounds, the magnetic interaction between adjacent layers of Mn ions, at room temperature, is ferromagnetic if the intralayer Mn-Mn bond length ($d^a_{Mn-Mn}$) > 2.85 Å, whereas it is antiferromagnetic if 2.84 Å< $d^a_{Mn-Mn}$ <2.85 Å. For $d^a_{Mn-Mn}$ <2.84, there is effectively no intra-layer spin component. Mn sublattice forms a tetragonal structure and the Mn-Mn inter-layer distance along the c-axis ($d^c_{Mn-Mn}$) is about 5.44 Å[16]. Since the magnetic ordering of the Mn sublattice is very much dependent on the lattice parameters, the overall magnetic state of these compounds is quite sensitive to temperature, owing to the thermal variation of the lattice parameters. Therefore, magnetism, magneto-thermal and magneto-transport properties are expected to be quite interesting in these compounds.

As part of our work on this series of compounds, recently we have reported the magnetic, magnetocaloric and magnetoresistance properties of ternary and pseudo-ternary compounds containing Sm and Gd, with the nonmagnetic elements Si and Ge[17,18]. It has been found that magnetic and related properties change considerably with Ge concentration in $GdMn_2Si_{2-x}Ge_x$ compounds[19]. Since Gd does not experience any considerable crystalline electric field (CEF), it is of interest to study the variations in the properties in systems such as $SmMn_2Si_{2-x}Ge_x$, in which the CEF would affect the properties. Furthermore, on account of the lattice parameter differences between Gd and Sm compounds, the magnetic and other related properties would differ. With this aim, in this paper, we report the effect of Ge substitution for Si on the magnetic, magnetocaloric and magnetoresistance properties of $SmMn_2Si_{2-x}Ge_x$ with x=0, 1 and 2.

## II. EXPERIMENTAL DETAILS

Polycrystalline samples of $SmMn_2Si_{2-x}Ge_x$ [ x=0, 1 and 2] were synthesized using the methods reported elsewhere[16-19]. The as-cast samples were characterized by room temperature power x-ray diffractograms (XRD), collected using Cu-K$_\alpha$ radiation. The magnetization (M) measurements, both under 'zero-field-cooled' (ZFC) and 'field-cooled' (FCW) conditions, in the temperature (T) range of 5-150 K and up to a maximum field (H) of 80 kOe were performed using a vibrating sample magnetometer (VSM, OXFORD instruments). In the ZFC mode, the samples were cooled in the absence of a field and the magnetization was measured during warming by applying a nominal field of 200 Oe. During the FCW measurement, the samples were cooled in presence of a field and the magnetization data was collected during warming, under the same field (200 Oe). The heat capacity (C) and electrical resistivity (ρ) were measured down to 2 K and in fields up to 50 kOe, using a physical property measurement system (PPMS, Quantum Design). The heat capacity was measured using the relaxation method and the electrical resistivity was measured by employing the linear four-probe technique. The magnetocaloric effect has been calculated in terms of isothermal magnetic entropy change using the M-H-T data.

**III. RESULTS AND DISCUSSION**

The Reitveld refinement of SmMn$_2$Si$_{2-x}$Ge$_x$ compounds at room temperature confirms that all the three compounds are single phase with the ThCr$_2$Si$_2$ structure (space group=I4/mmm). The lattice parameters and the Mn-Mn bond lengths along the *a*-axis and the *c*-axis ($d^a_{Mn-Mn}$ and $d^c_{Mn-Mn}$), as obtained from the refinement, are given in Table 1. It may be noticed from the table that both $d^a_{Mn-Mn}$ and $d^c_{Mn-Mn}$ increase with increase in the Ge content. This variation is attributed to the larger ionic radius of Ge compared to that of Si. Furthermore, it may be noticed from the table that $d^a_{Mn-Mn}$ for the compounds with x=0 and 1 is less than the critical value (2.84 Å) needed for the presence of intra-layer spin component of the Mn sublattice, whereas it is just above the critical value in the case of x=2. Table 1 shows the summary of the structural parameters in these compounds.

Table 1. Lattice parameters (*a* & *c*), unit cell volume (*V*), Mn-Mn bond lengths in SmMn$_2$Si$_{2-x}$Ge$_x$ compounds

| x | a (Å) | c (Å) | V (Å$^3$) | $d^a_{Mn-Mn}$ (Å) | $d^c_{Mn-Mn}$ (Å) |
|---|---|---|---|---|---|
| 0 | 3.973 | 10.503 | 165.812 | 2.808 | 5.252 |
| 1 | 4.013 | 10.701 | 172.300 | 2.811 | 5.350 |
| 2 | 4.062 | 10.889 | 179.672 | 2.872 | 5.445 |

Fig. 1 shows the temperature dependence of the magnetization data of $SmMn_2Si_{2-x}Ge_x$ compounds in an applied field of 200 Oe, under ZFC and FCW conditions. As can be seen, the compounds with x=0 and 1 show identical M-T behavior. The sharp decrease in the magnetization seen in both these compounds is attributed to the magnetic order-disorder transition occurring in the Sm sublattice. We designate the temperature of this transition as $T_C^{Sm}$, which is calculated from the d$M$/d$T$ vs. T plot. The $T_C^{Sm}$ for the compounds with x= 0, 1 and 2 are found to be 39 K, 37 K and 110 K, respectively. In the compound with x=2, there is an additional magnetic transition from a low moment state to a high moment state at T*. It has been reported that the magnetic state of $SmMn_2Ge_2$ is ferromagnetic for $T < 110$ K and $146 < T < 350$ K, antiferromagnetic for $110K < T < 146$ K and paramagnetic for $T > 350$ K[16,17]. It is to be noted that the magnetic transition temperatures observed in the compounds with x=0 and x=2 are in good agreement with the reported values. It should be noted that while $GdMn_2Si_{2-x}Ge_x$ is antiferromagnetic at all temperatures above $T_C^{Gd}$, $SmMn_2Si_{2-x}Ge_x$ possesses a ferromagnetic phase in certain temperature regime above $T_C^{Sm}$. In addition to the transitions mentioned, in $SmMn_2Ge_2$, another transition can be seen at about 30 K. The same observation has been reported other authors[12] as well and the reason for this transition is not exactly known. The $T_C^{Sm}$ and T* values obtained in $SmMn_2Si_{2-x}Ge_x$ are listed in Table 2.

It is interesting to note that the Sm ordering temperature in this series is almost constant for x= 0 and 1, whereas it shows considerable increase for x=2. It may be recalled here that $d_{Mn-Mn}^a$ for the compounds with x= 0 and 1 is less than the critical value required for the occurrence of intra-layer spin component whereas it is larger than the critical value in the compound with x=2. Therefore, based on the lattice parameter dependence of the magnetic ordering of the Mn sublattice mentioned earlier, the compound with x=2 would have a nonzero intra-layer Mn spin component, in addition to the interlayer component. Consequently, there will be an overall increase in the Mn-Mn and Sm-Mn exchange strengths and hence the $T_C^{Sm}$ increases with $x$. A similar observation has been made in $GdMn_2Si_{2-x}Ge_x$ compounds as well [19].

Fig. 2 shows the magnetization isotherms of SmMn$_2$Si$_{2-x}$Ge$_x$ compounds at 5 K. It can be seen from the figure that the magnetic moment, for an applied field of 80 kOe, of the compounds with x= 0 and 1 is about 1.1 μ$_B$/f.u. and 1.4 μ$_B$/f.u., respectively, whereas it is 2.6 μ$_B$/f.u. for the compound with x= 2. It may be mentioned here that in the RMn$_2$X$_2$ compounds, below the rare earth ordering temperature, the Mn moments order ferromagnetically for light rare earth. Therefore, the increase in the magnetization value for the compound with x=2 suggests an increase in the Mn moment as x is increased from 0 to 2. This observation is consistent with the increase in $T_C^{Sm}$ values with x, as mentioned earlier.

Fig. 3 shows the magnetization isotherms of SmMn$_2$Ge$_2$ at 75, 130 and 160 K. These three temperatures correspond to the regimes T<$T_C^{Sm}$, $T_C^{Sm}$<T<T* and T>T*. The inset shows the low field region of the isotherms at 120 K, which lies between $T_C^{Sm}$ and T*. A slow increase in *M* is consistent with the low-field antiferromagnetic (AFM) state in this compound. As is clear from the inset, the magnetization shows several jumps for fields less than about 8 kOe. At higher fields, saturation-like behavior indicates that the compound is ferromagnetic (FM). The observed jumps may be attributed to the micro-level inhomogeneities in the polycrystalline sample as well as the large anisotropy. A similar observation has been earlier made on polycrystalline DyNi[20]. The critical fields for the onset and completion of metamagnetic transition at various temperatures are shown in Fig. 4. In this figure, FM1 and FM2 represent the two ferromagnetic phases in the temperature range of T<110 K and 146<T<350 K, respectively.

Table 2 Values of T$_C^{Sm}$, T*, electronic heat capacity coefficient, density of states at the Fermi level and the maximum isothermal magnetic entropy change in SmMn$_2$Si$_{2-x}$Ge$_x$ compounds.

| x | $T_C^{Sm}$ (K) | T* (K) | γ(mJ/mol K) | N(E$_F$) (states/eV atom) | (-ΔS$_M$)max (J/kg K) at $T_C^{Sm}$ |
|---|---|---|---|---|---|
| 0 | 39 | -- | 5.84 | 0.50 | 3.4 |

| 1 | 37  | --  | 14.0 | 1.19 | 4.5  |
| 2 | 110 | 145 | 24.2 | 2.10 | -0.1 |

Fig.5a-c shows the temperature variation of heat capacity of $SmMn_2Si_{2-x}Ge_x$ compounds in zero field. Generally, the heat capacity C (T) for a magnetic material can be described as the sum of phonon, electronic and magnetic contributions. i.e.

$$C(T) = C_{ph}(T) + C_e(T) + C_M(T) \quad\quad\quad (1)$$

We have fitted the C-T data in the temperature range 2-15 K, with the usual expression of the heat capacity of metallic systems at low temperature, i.e.,

$$\frac{C}{T} = \gamma + \beta T^2 \quad\quad\quad (2),$$

Where γ (coefficient of electronic heat capacity) is given by

$$\gamma = \frac{k_B^2 \pi^2}{3} N(E_F)(1+\lambda) \quad\quad\quad (3)$$

Here N ($E_F$) is the electronic density of states of the bare electrons at the Fermi level and (1+λ) is the mass enhancement factor caused by electron- phonon interaction[21]. Here $k_B$ is the Boltzmann constant. The coefficient β in equation 2 is given by

$$\beta = \frac{12\pi^4 R}{5\theta_D^3} \cong \frac{1944}{\theta_D^3} \quad\quad\quad (4).$$

Where R is the universal gas constant and $\theta_D$ is the Debye temperature. Using the equations 2 and 3, we have calculated N($E_F$) for all the three compounds. In the case of $SmMn_2Si_2$ and $SmMn_2Ge_2$, the values are in good agreement with the previous results of theoretical electronic band calculations[21]. It is observed that the compounds with ferromagnetic Mn-Mn interlayer coupling possess larger N($E_F$) values than those of the compounds with antiferromagnetic Mn-Mn interalayer coupling [21]. On substituting Ge for Si, like $d_{Mn-Mn}^a$, γ and N($E_F$) increase, as is evident from Tables 1 and 2. This indicates a good correlation between the lattice parameters and the electronic structure in theses compounds.

We have also calculated the magnetic contribution to the total heat capacity. This has been done by subtracting the nonmagnetic contributions given by equation 5 from the experimental data.

$$C_{total} = C_e(T) + C_{ph}(T) = \gamma T + 9NR(T/\theta_D)^3 \int_0^{\theta_D/T} \frac{x^4 e^x}{(e^x-1)^2} dx \quad \ldots\ldots (5)$$

Where the second term corresponds to the phonon contribution. Here N is the number of atoms per formula unit (N=5 in this case). It has been found that the values of $\theta_D$= 341K and $\gamma$=24.2 for $SmMn_2Ge_2$ agree well with the reported values of $\theta_D$= 341K and $\gamma$=30.3 mJ/mol K respectively[22]. By using the relation $\theta_D \alpha\ M^{-1/2}$, we have calculated the $\theta_D$ values to be 386 K for $SmMn_2Si_2$. Fig. 5 shows the temperature dependence of $C_M$ for all the three compounds. The solid line represents the theoretically calculated electronic and phonon contributions. The magnetic contribution is represented by filled squares. The inset shows the magnetic entropy variation. The value of $C_M$ at $T_C^{Sm}$ is about 17.7 J/mol K in $SmMn_2Si_2$ and 12 J/mol K in $SmMn_2Ge_2$. In $SmMn_2Ge_2$ another sharp peak at 150 K with a magnitude of 17.3 J/mol K is observed. In both the compounds, $C_M$ increases with increasing temperature. As can be seen from the inset, at $T_C^{Sm}$ the values of magnetic entropy are 8.9 J/mol K for x=0 and 13 J/mol K for x= 2. It is of interest to note that the Rln (2J+1) value for $Sm^{3+}$ (J=5/2) ion is 14.8 J/mol K. In all the three compounds, at high temperatures, the $S_M$ versus T plot shows a convex curvature, which must be due to the Mn sublattice contribution.

All the magnetic transitions, except the one at 30 K in $SmMn_2Ge_2$, appear as peaks in the $C_M$ vs. T plots. In addition, another peak is observed at low temperatures, which is attributed to the Schottky effect. The magnetic contribution was fitted by taking into account the Schottky contribution[23], i.e.

$$C_M = \frac{R}{T^2} \left( \frac{\sum_{i=1}^{6} E_i^2 \exp(-\frac{E_i}{k_B T})}{\sum_{i=1}^{6} \exp(-\frac{E_i}{k_B T})} - \left(\frac{\sum_{i=1}^{6} E_i \exp(-\frac{E_i}{k_B T})}{\sum_{i=1}^{6} \exp(-\frac{E_i}{k_B T})}\right)^2 \right) \ldots (6)$$

According to the Kramer's theorem, for odd electron systems there is always a two-fold degeneracy which cannot be removed by any crystalline electric field. Therefore, in the case of $Sm^{3+}$(J=5/2), the ground state multiplet splits into three doublets, in the paramagnetic state. In the magnetically ordered state (T<$T_C^{Sm}$), one expects a maximum of six singlet states. Fig. 6a-b shows the $C_M$ vs. T data fitted to the equation 6, in the range T<$T_C^{Sm}$. The corresponding energy levels of the ground state multiplet are 0 K, 63.7 K, 104.3 K(doublet), 128.6 K and 137.4 K in $SmMn_2Si_2$ and 0 K , 84.9 K, 140.7 K, 153.3 K, 183.8 K and 223.5 K in $SmMn_2Ge_2$ respectively. The level scheme obtained by the fitting is shown in each plot.

The spin wave contribution to the heat capacity of $SmMn_2Si_{2-x}Ge_x$ compounds at low temperatures is also analyzed. Fig.7 shows the temperature variation of heat capacity of all the three compounds below the $T_C^{Sm}$, fitted to the relation $C = a\exp(-\frac{\Delta}{T})(\frac{\Delta}{T}+2+\frac{2T}{\Delta})$, where $\Delta$ is the minimum energy required to excite a spin wave in the anisotropy field[24]. As shown in the figure, the value of $\Delta$ is the same for x=0 and 1, whereas it increases considerably in x=2. A similar variation has been observed in $GdMn_2Si_{2-x}Ge_x$ as well [19]. The reason for this increase in the case of x=2 is that the magneto crystalline anisotropy increases with increase in Ge concentration.

The magnetocaloric effect in these compounds has been measured in terms of the isothermal magnetic entropy change ($\Delta S_M$) for various temperatures and applied magnetic fields. The $\Delta S_M$ is calculated from magnetization isotherms $M(T_i, H)$, obtained at different temperatures $T_i$ close to $T_C^{Sm}$, using the Maxwell's relation[25,26].

$$\Delta S_M(T_i, H_2) = \int_{H_1=0}^{H_2} \left(\frac{\partial M(T,H)}{\partial T}\right)_{T_{av,i}} dH$$

$$\approx \frac{1}{T_{i+1}-T_i}\int_0^H [M(T_{i+1},H) - M(T_i,H)]dH$$

Here, $T_{av, i} = (T_{i+1}+T_i)/2$ is the average temperature and $\Delta T = T_{i+1}-T_i$ is the temperature difference between the magnetization isotherms measured at $T_{i+1}$ and $T_i$, when the magnetic field is changed from $H_1=0$ to $H_2$.

Fig. 8a-c shows the temperature variation of $\Delta S_M$ for a field change ($\Delta H$) of 20kOe and 50kOe, for all compounds. It can be seen from Fig. 8a and b that $\Delta S_M$ vs. T plots for SmMn$_2$Si$_2$ and SmMn$_2$SiGe show a broad maximum near $T_C^{Sm}$, unlike the general observation of a sharp peak in many materials. Broad maximum and symmetrical nature of the peak near $T_C^{Sm}$ indicate second order transition in these two compounds. In the case of SmMn$_2$Ge$_2$, as is evident from Fig. 8c, we find that $\Delta S_M$ ($T_C^{Sm}$) = 0.8 J/kg K and $\Delta S_M$ (T*) =1.3 J/kg K for a field change of 1 kOe. The absolute values of $\Delta S_M$ are much smaller than 14 J/kg K (i.e. $R$ln2) for a simple spin state ($S$ = 1/2) or 37 J/kg K (i.e. $R$ln6) for $J$ = 5/2. Furthermore, $\Delta S_M(T_C^{Sm})$ values are found to decreases with increasing field, as shown in the fig. 5c. This may be due to the admixture and crossing of the CEF levels[27]. A Similar result has been observed in PrNi$_5$ [28,29].

Fig. 9a-b shows the temperature variation of electrical resistivity of SmMn$_2$SiGe and SmMn$_2$Ge$_2$ compounds. As can be seen, the resistivity values increases with increase in Ge concentration. The increases in the magnitude of resistivity may be partly due to the extra Mn moment, as compared to that in SmMn$_2$Si$_2$. Two anomalies, at 110K and 146K, corresponding to the magnetic transitions can be seen in the case of SmMn$_2$Ge$_2$.

The general variation of resistivity with temperature has been found to show a good fit to the relation $\rho \propto T^n$ with $n$ in the range of 1.5-1.6, except the region corresponding to the antiferromagnetic state[30]. Value of n close to 1.5 indicates the ferro/ferri magnetic nature of the compound. The excess resistivity in the antiferromagnetic state compared to the ferromagnetic state ($\Delta\rho = \rho_{total} - \rho_{f\,erro}$) in the case of SmMn$_2$Ge$_2$ is given in inset of the figure10b. The result indicates that the resistivity in antiferromagnetic region is about 20% more than that in the ferromagnetic region. The higher resistivity in the antiferromagnetic state than that in the ferromagnetic state is in good agreement with the behavior of polycrystalline Gd$_{0.975}$La$_{0.025}$Mn$_2$Si$_2$ compound. Fujitwara et. al[30] have reported the variation of resistivity with temperature, along the c-axis ($\rho^{axis}$) and in the c-plane ($\rho^{plane}$) in single crystals of Gd$_{0.975}$La$_{0.025}$Mn$_2$Si$_2$. They have shown that the variation of $\rho^{axis}$ obeys $T^{1.5}$ behavior, while $\rho^{plane}$ shows a $T^{1.8}$ dependence. Furthermore, they have shown that the resistivity in the *c* plane is smaller than that along the c axis, suggesting that the mobility of Mn conduction electrons is larger in the c-plane than along the c- axis. This reflects the two dimensional (layered) arrangement of Mn atoms. Lattice parameter variation in $SmMn_2Si_{2-x}Ge_x$ indeed shows that the *c* lattice parameter increases with x, indicating that the two-dimensional nature is more prominent in $SmMn_2Ge_2$, as compared to the other two compounds.

## IV. CONCLUSION

We find that the change in the magnetic properties as Si is gradually substituted with Ge arises mainly from the variation in the lattice parameters. Variation of both the Sm ordering temperature as well as the saturation magnetization reflect this variation. Multiple magnetic transitions and re-entrant ferromagnetism are observed in $SmMn_2Ge_2$. Magnetization data also show another magnetic transition in the re-entrant phase (at about 30 K), which could not be explained. The density of states at the Fermi level is also found to be quite high in $SmMn_2Ge_2$, as compared to the other two compounds of the series. The magnetic contributions to the heat capacity shows Schottky peak, in addition to the magnetic transitions. The energy level scheme corresponding to the ground state multiplet has been obtained. The temperature variation of isothermal magnetic entropy change is quite anomalous in $SmMn_2Ge_2$. The electrical resistivity increases with Ge concentration. The excess resistivity in the antiferromagnetic region has been calculated.

Acknowledgements

One of the authors (KGS) acknowledges the financial assistance received from ISRO, Govt. of India for carrying out this work.

Figure captions

Fig. 1 Temperature variation of magnetization of SmMn$_2$Si$_{2-x}$Ge$_x$ compounds in a field of 200 Oe under ZFC and FCW conditions.

Fig. 2. M-H isotherms of SmMn$_2$Si$_{2-x}$Ge$_x$ compounds at 5 K.

Fig. 3. M-H isotherms of SmMn$_2$Ge$_2$ at 75 K, 125 K and 160 K. The insets show the expanded low field region of the M-H plot at 120 K.

Fig. 4. Temperature variation of critical fields in antiferromagnetic region in SmMn$_2$Ge$_2$

Fig.5 Temperature variation of heat capacity of (a) SmMn$_2$Si$_2$ (b) SmMn$_2$SiGe and (c)SmMn$_2$Ge$_2$ in zero applied field (C$_{exp}$). The solid line represents the theoretically calculated electronic and phonon contributions (C$_{Th}$) . The filled squares represent the magnetic contribution (C$_M$). The inset shows the magnetic entropy variation.

Fig.6 Schottky fitting of the heat capacity in (a) SmMn$_2$Si$_2$ and (b) SmMn$_2$Ge$_2$ compounds. Open circle and solid line show the experimental and theoretically fitted data respectively.

Fig.7 Variation of heat capacity in SmMn$_2$Si$_{2-x}$Ge$_x$ compounds in the low temperature regime. Δ represents the energy gap in the spin wave spectrum.

Fig.8a-c Variation of isothermal magnetic entropy change of SmMn$_2$Si$_{2-x}$Ge$_x$ compounds with temperature for various field changes.

Fig.9 Temperature variation of resistivity (open circle) of (a) SmMn$_2$SiGe and (b) SmMn$_2$Ge$_2$ compounds. Solid line indicates the resistivity in the ferromagnetic state fitted

to $T^{1.5}$ law. Inset shows the excess resistivity ($\Delta\rho = \rho_{total} - \rho_{f\,erro}$) in the antiferromagnetic state in SmMn$_2$Ge$_2$

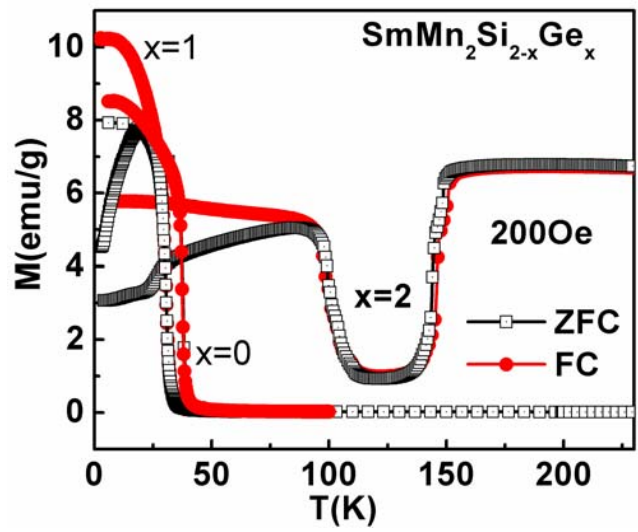

Fig.1

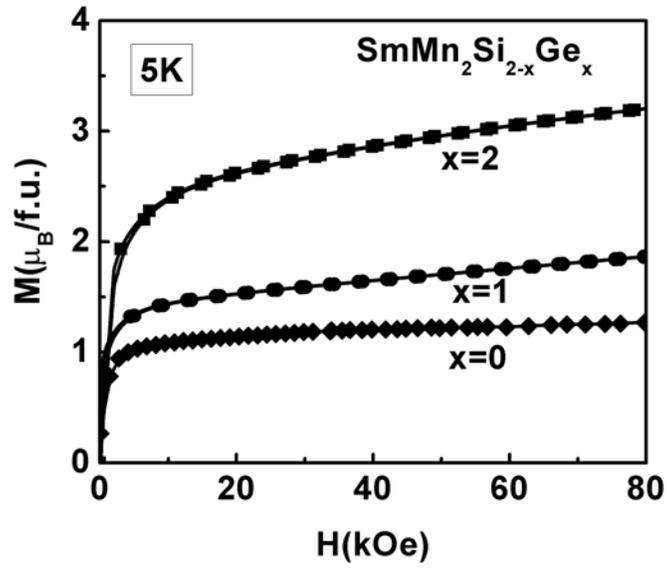

Fig.2

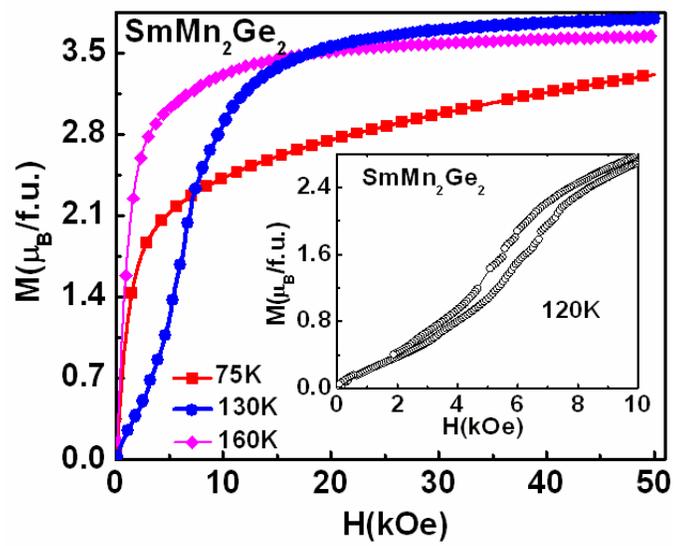

Fig.3

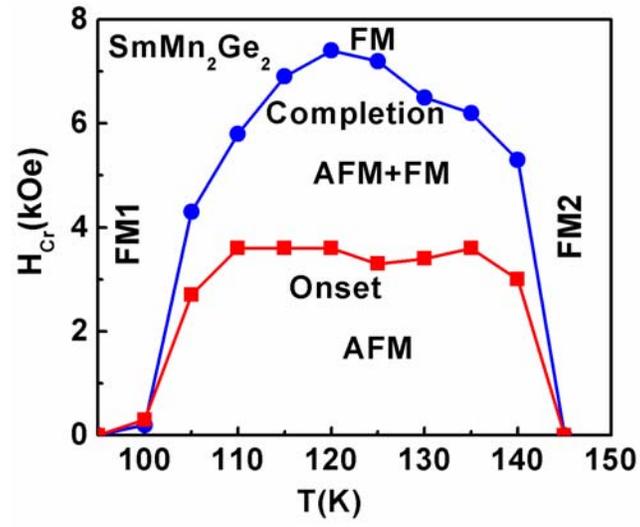

Fig.4

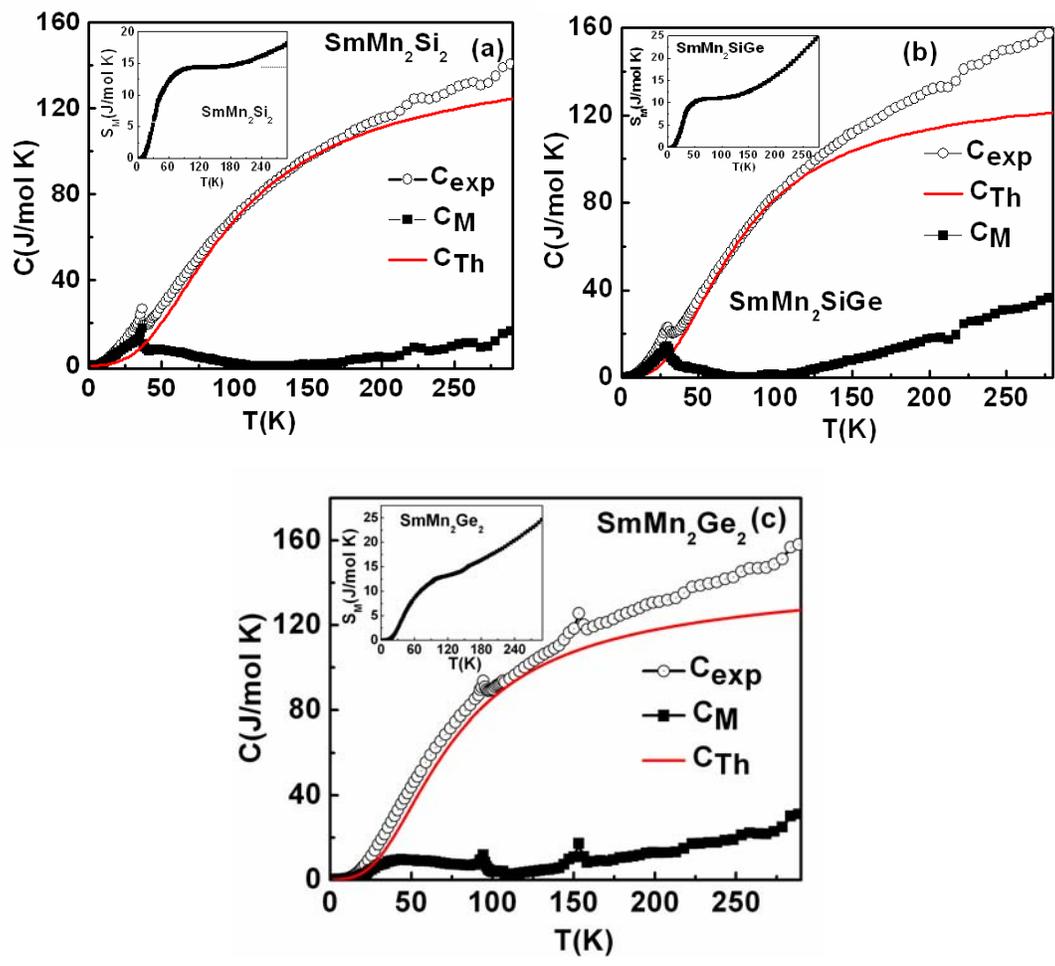

Fig. 5

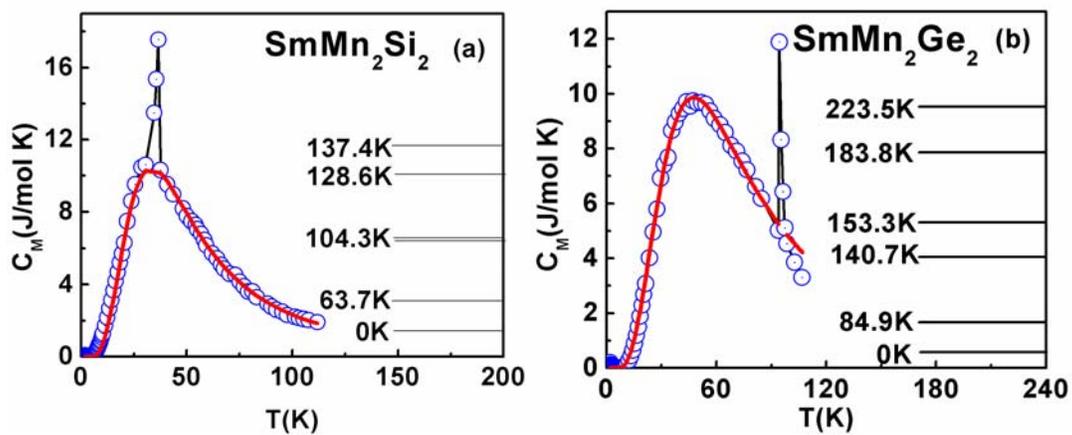

Fig.6

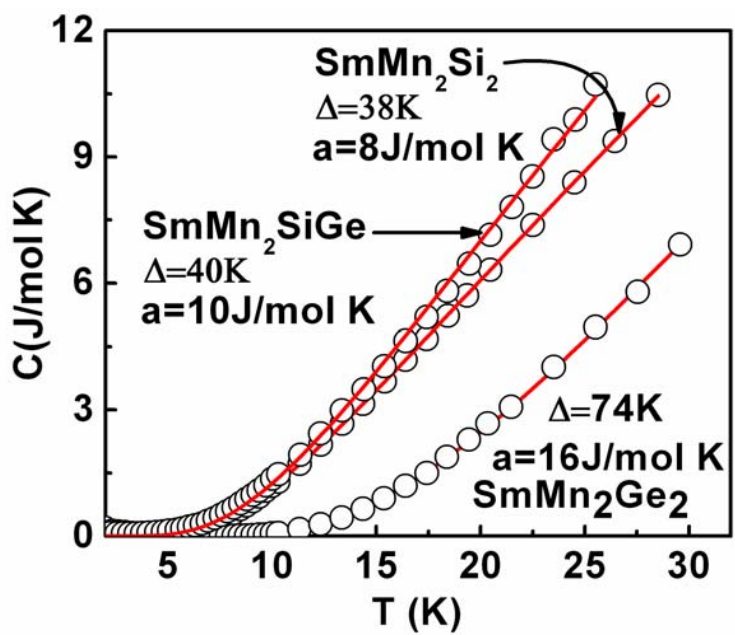

Fig.7

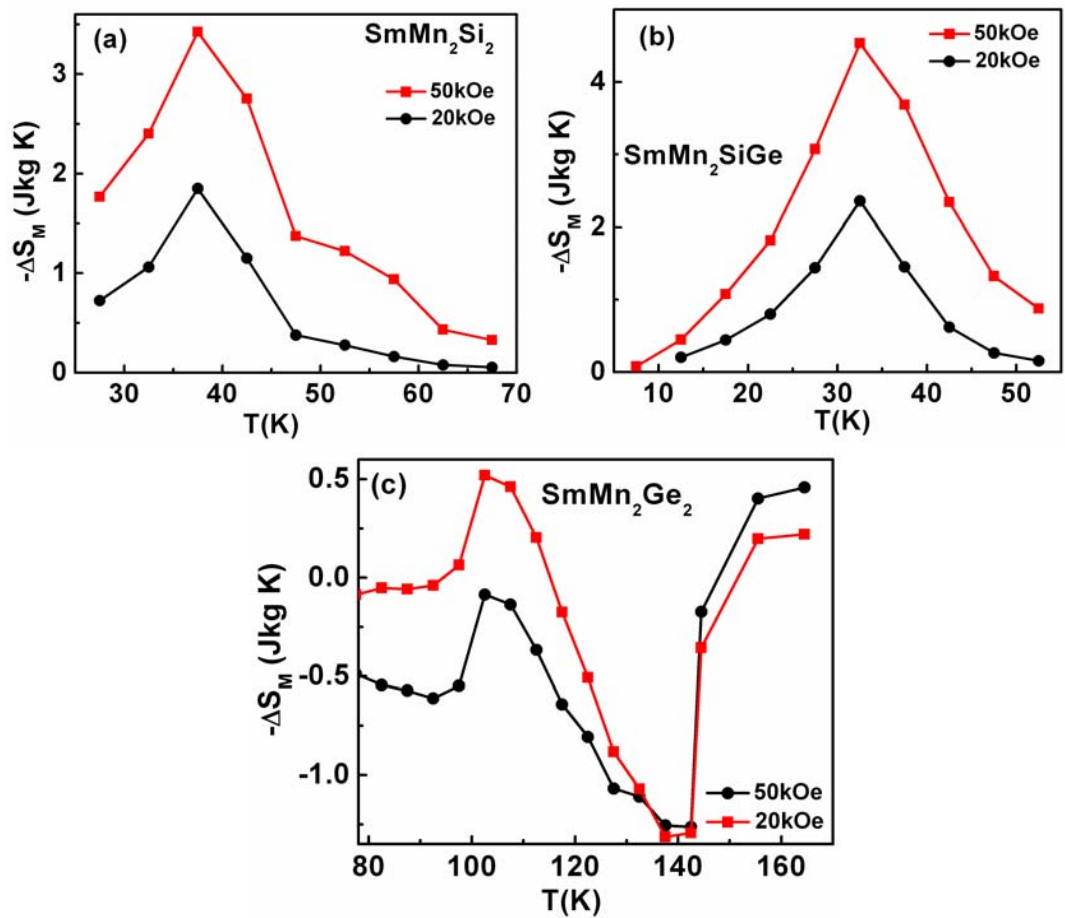

Fig.8

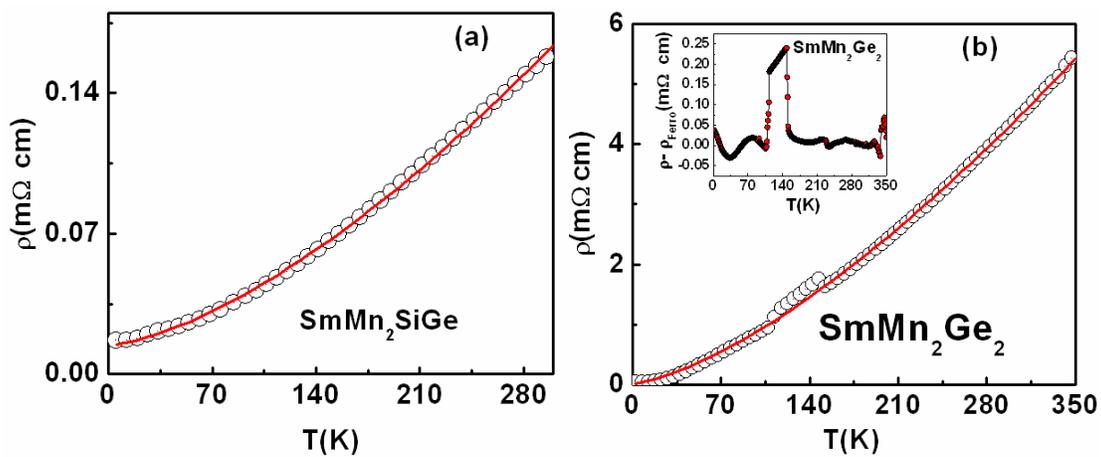

Fig. 9

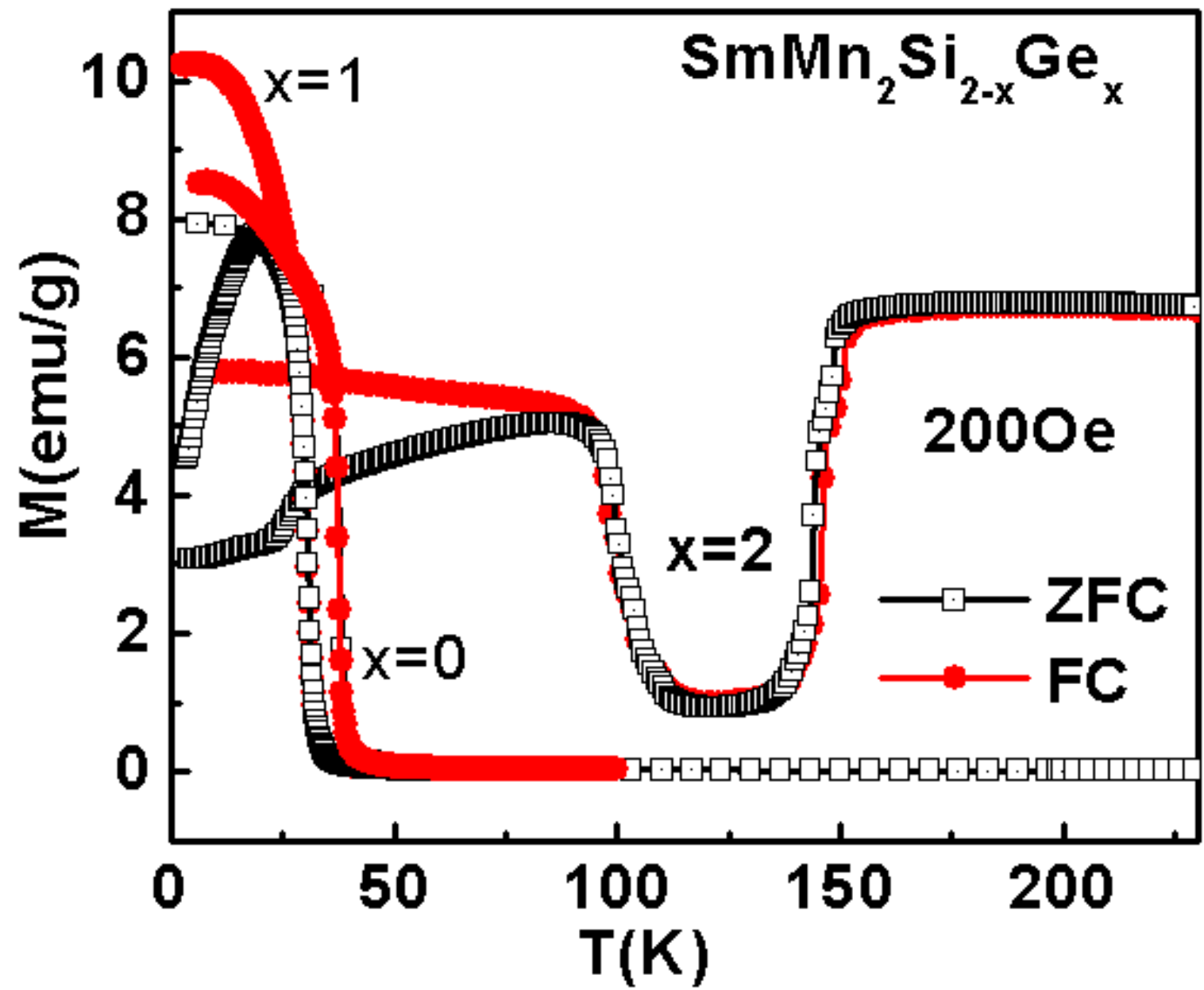

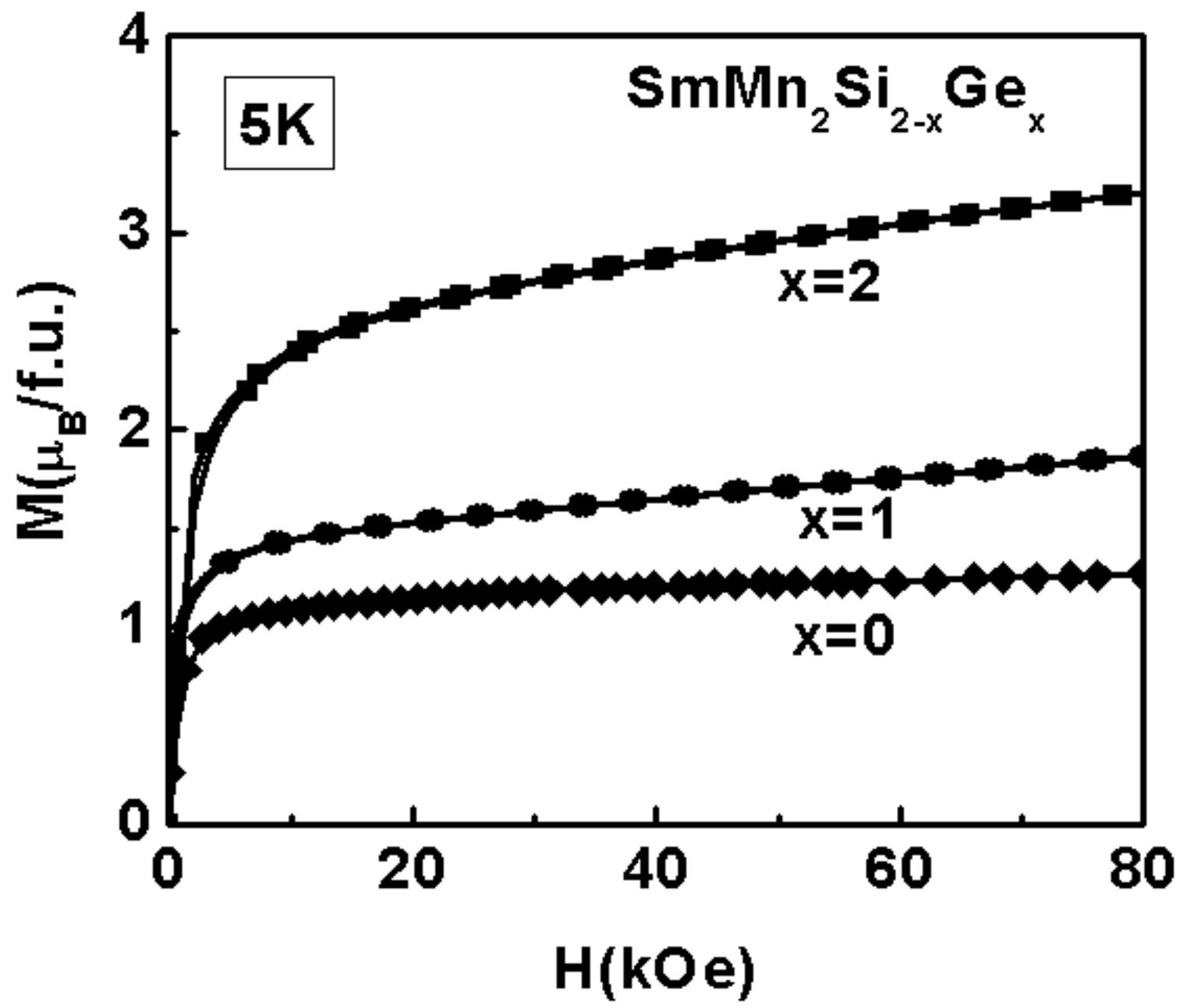

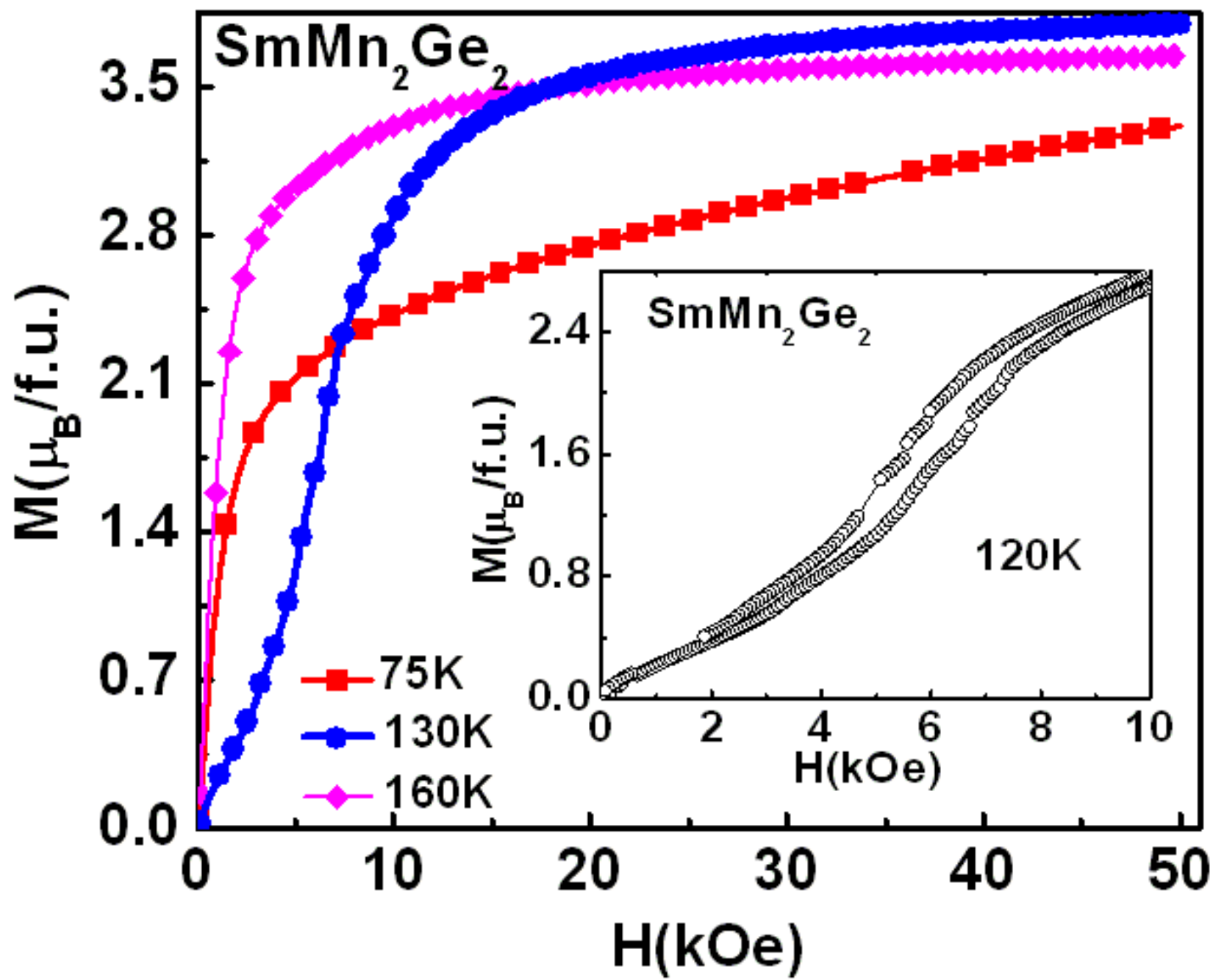

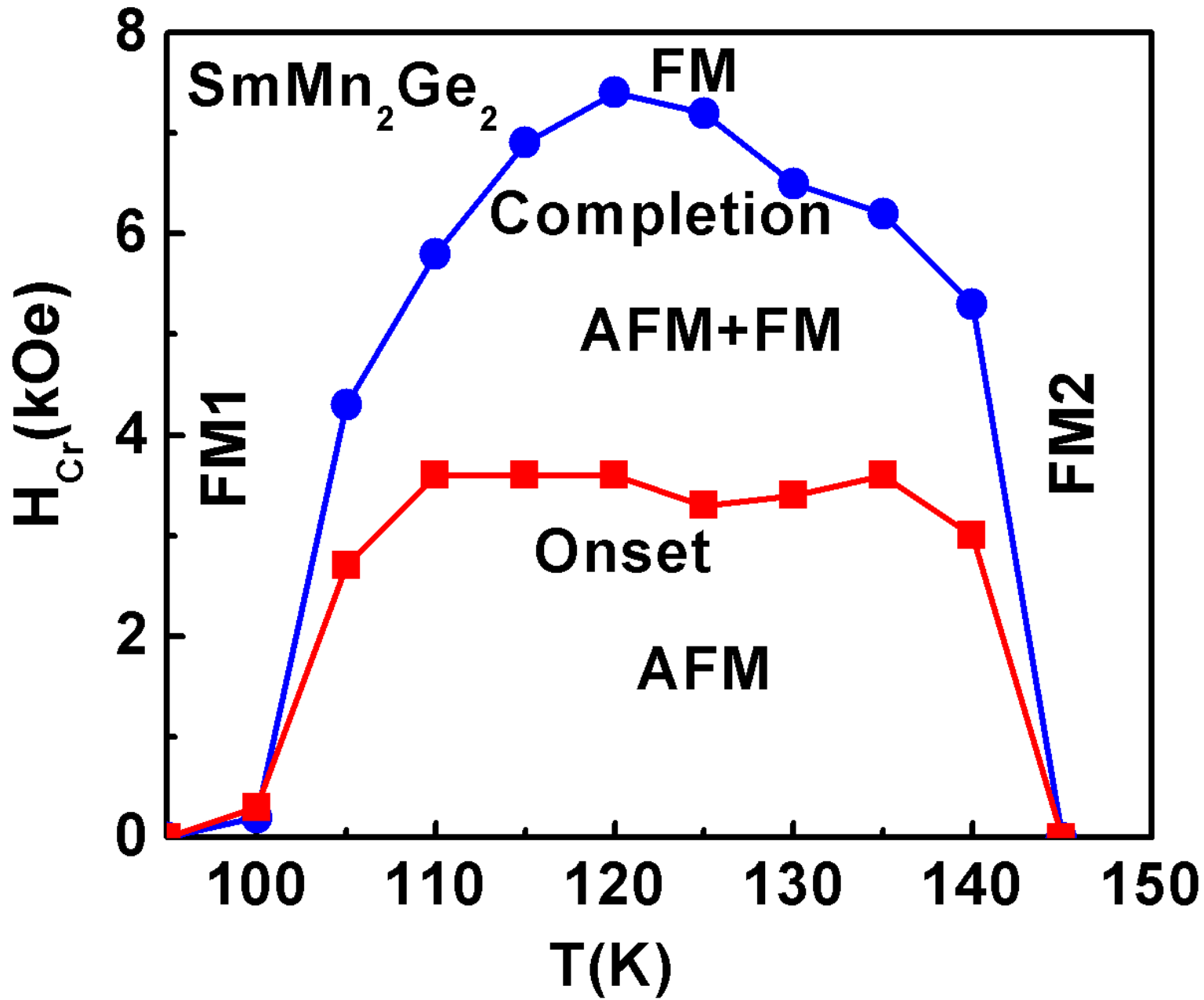

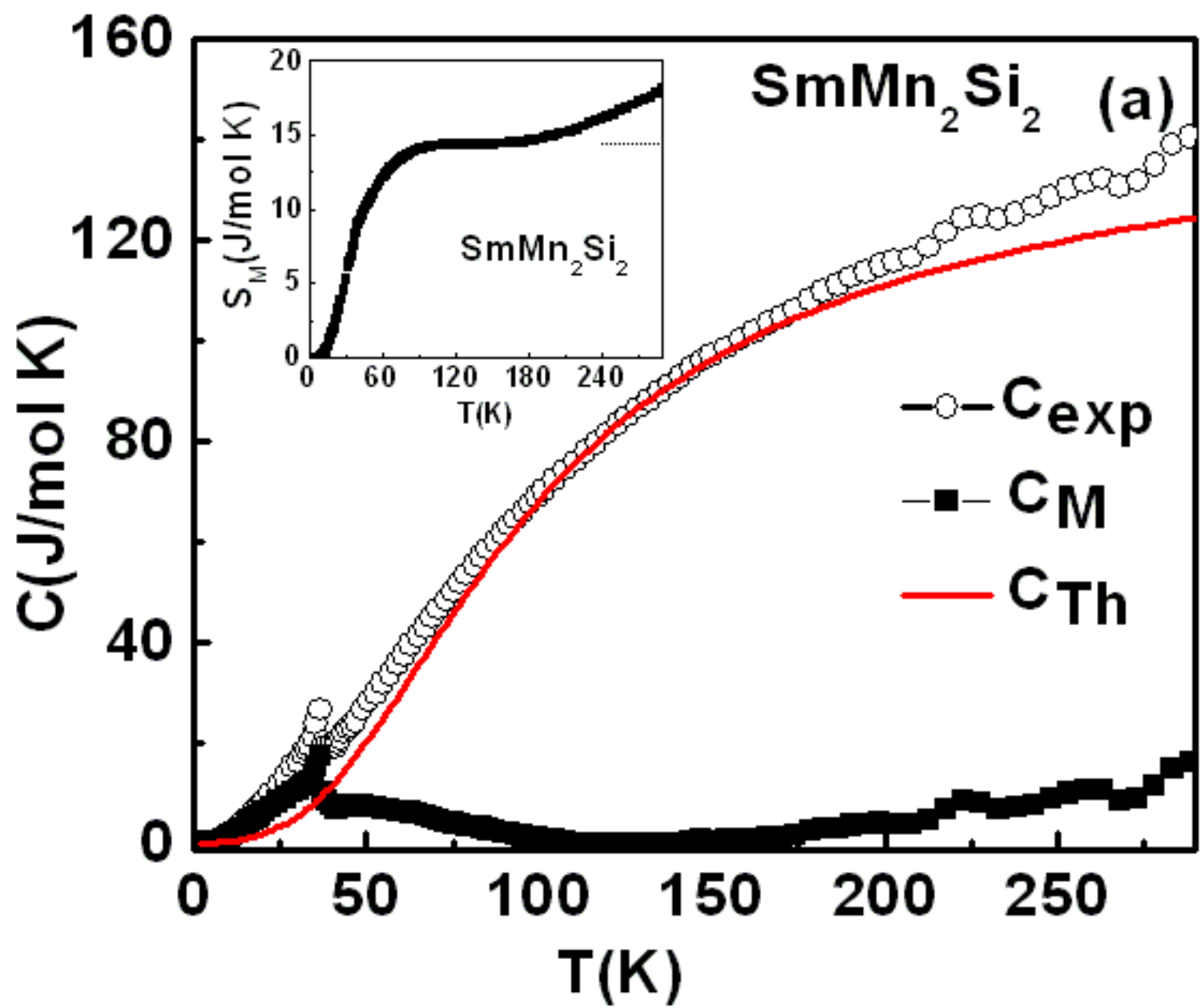

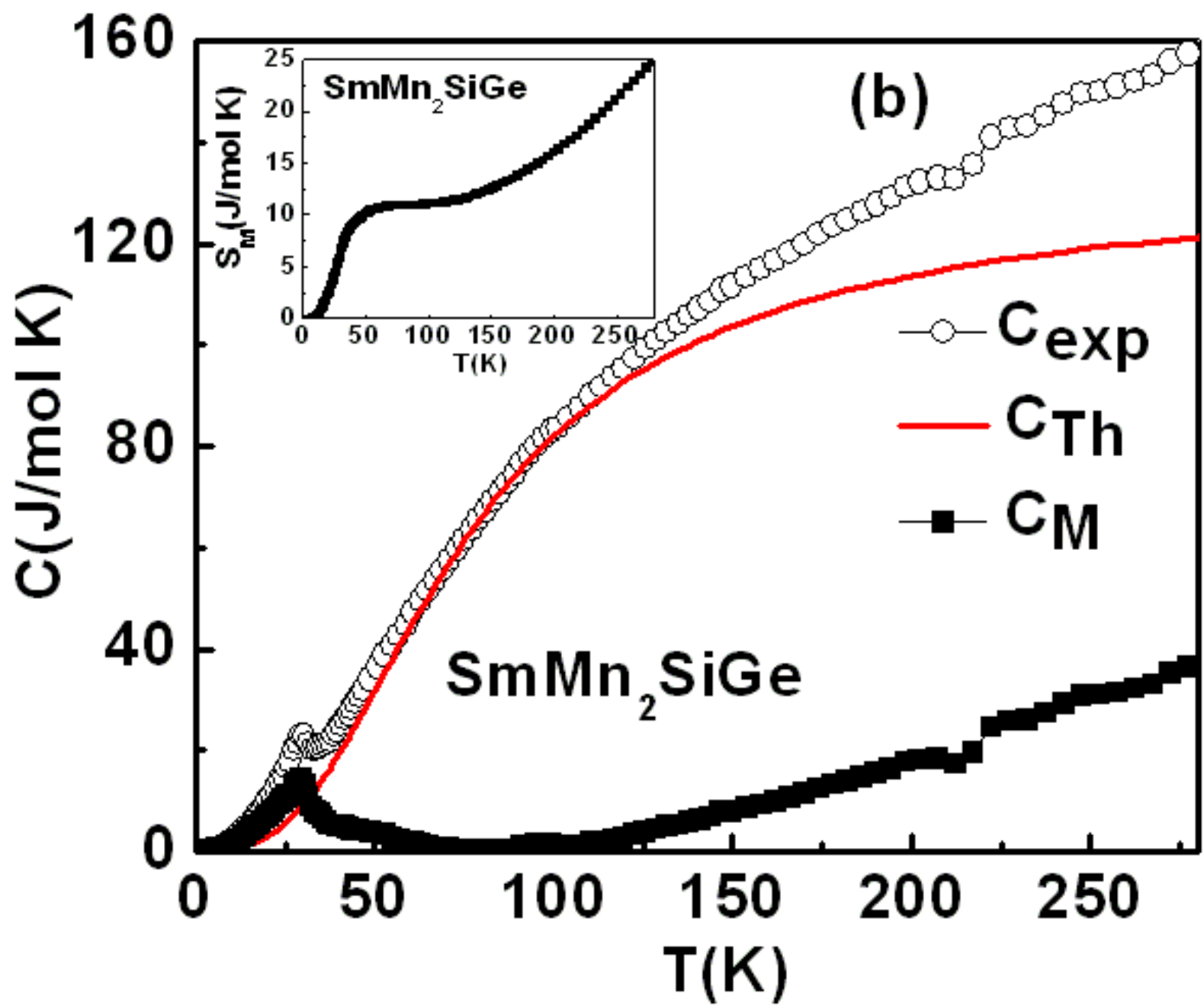

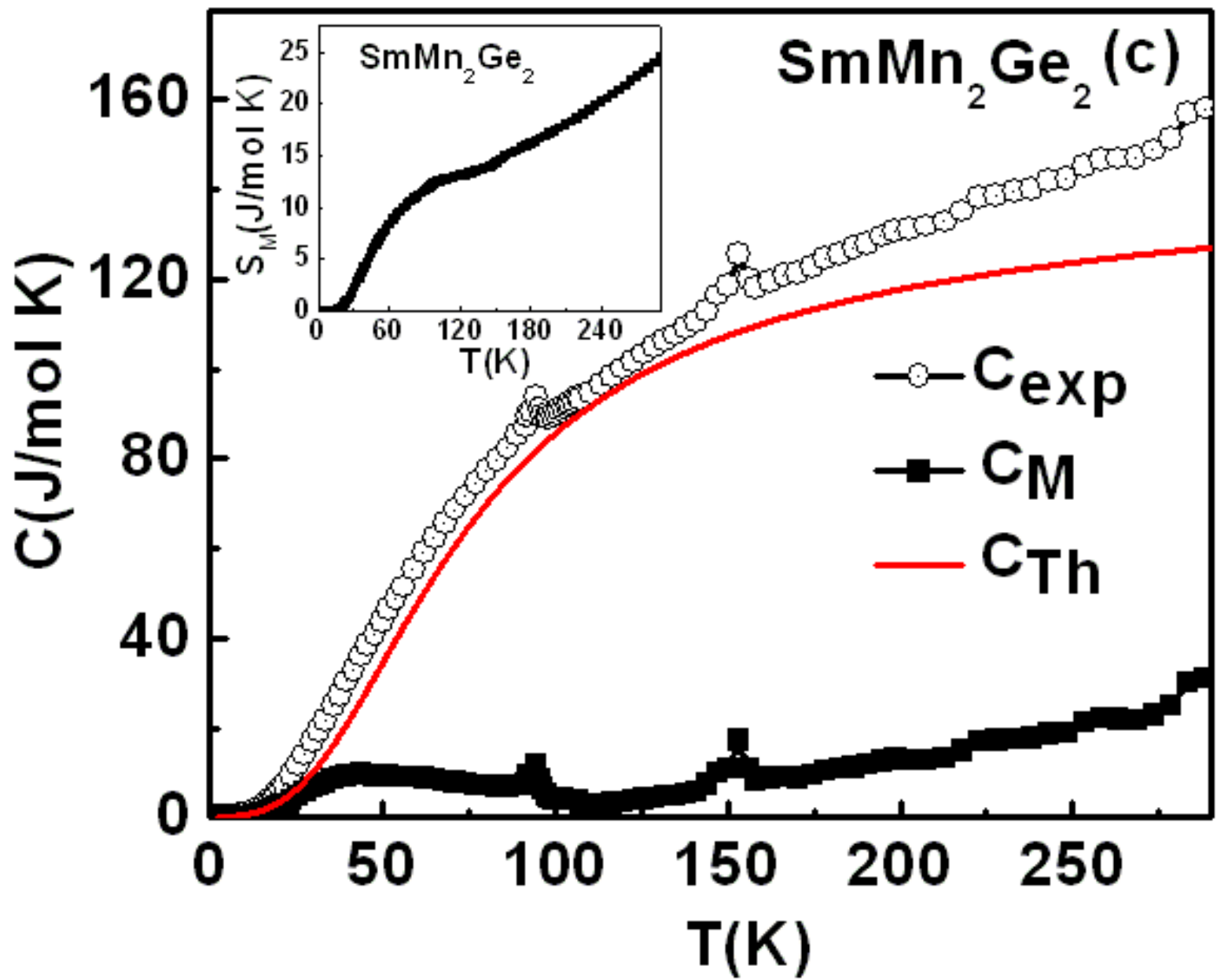

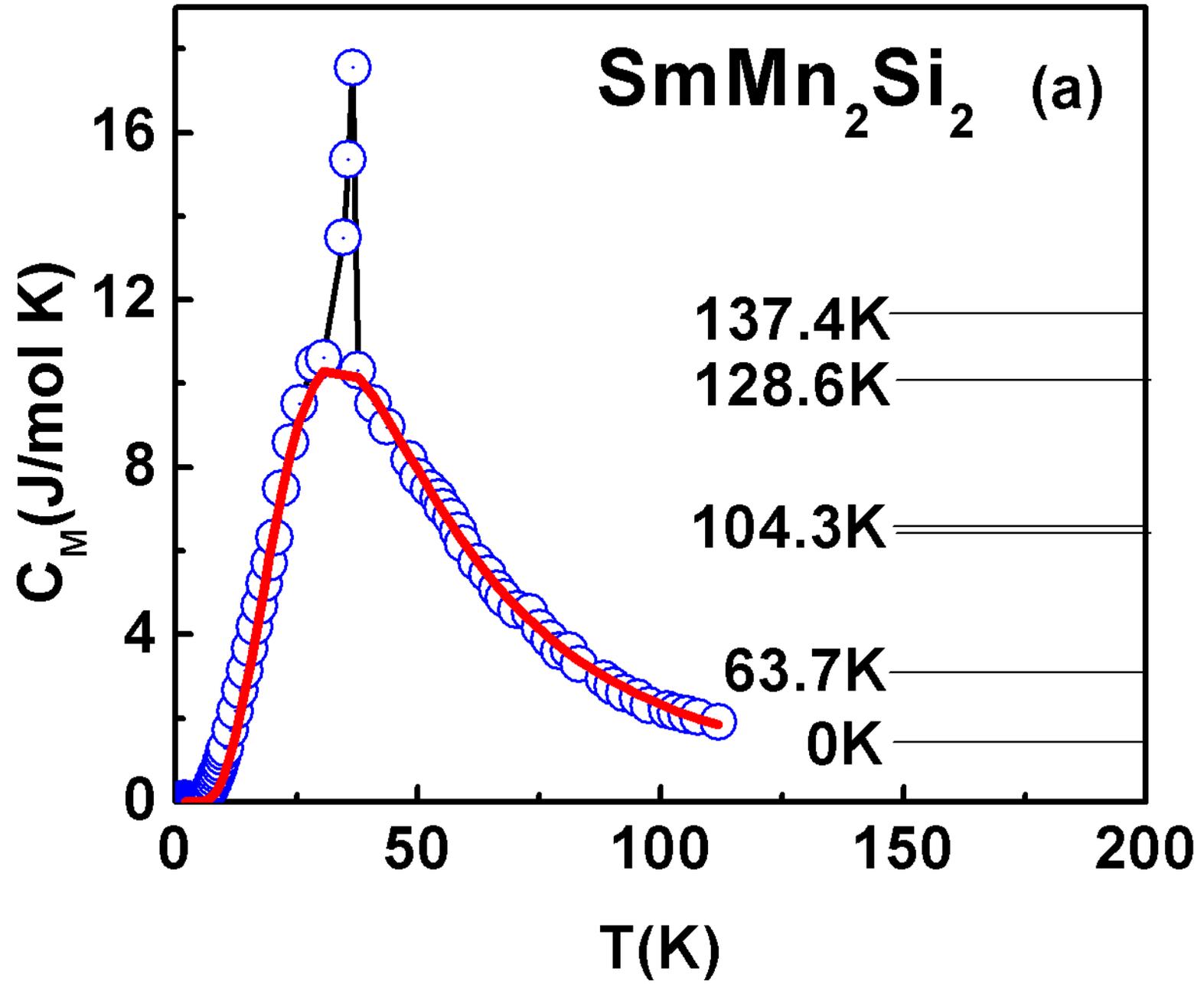

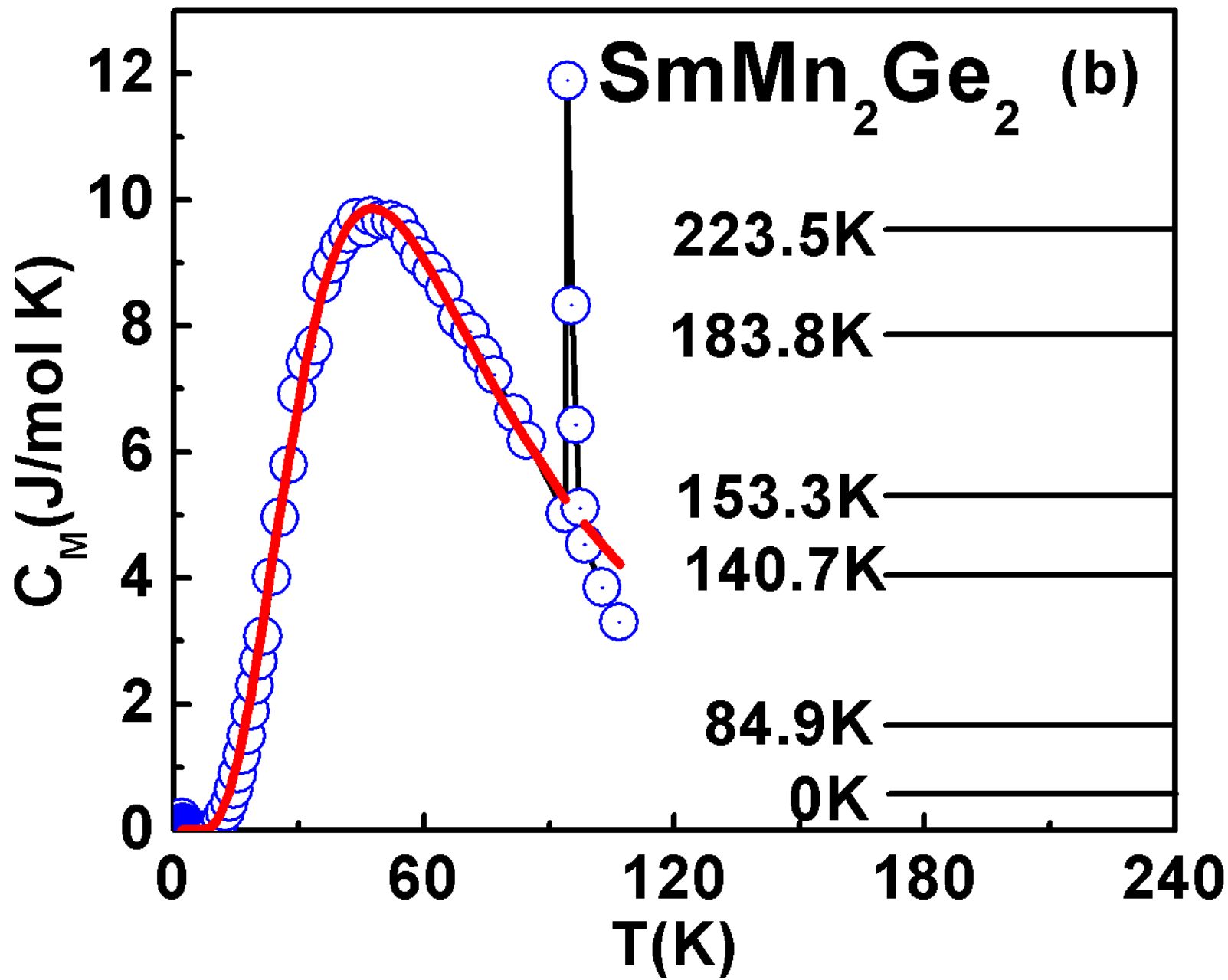

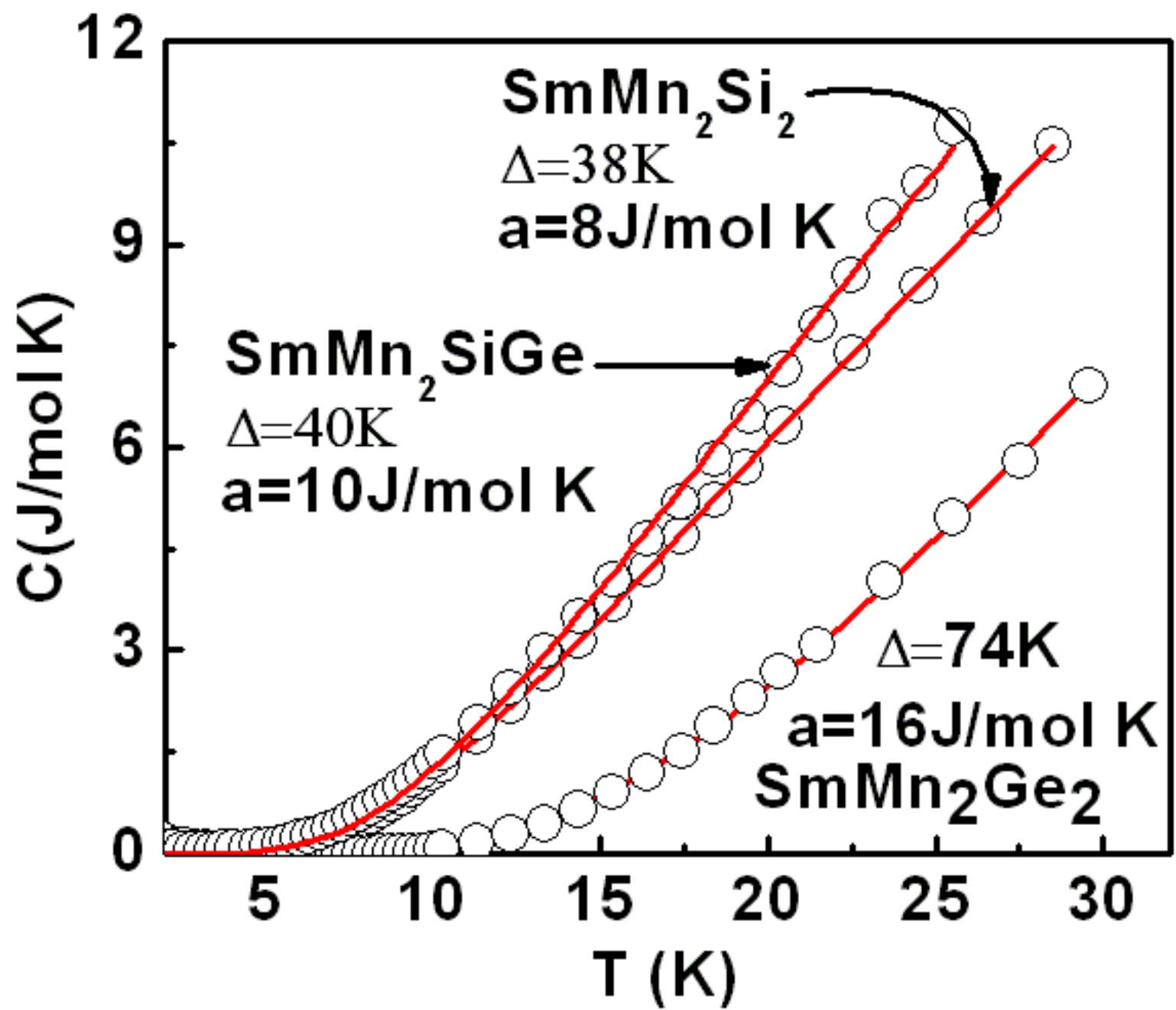

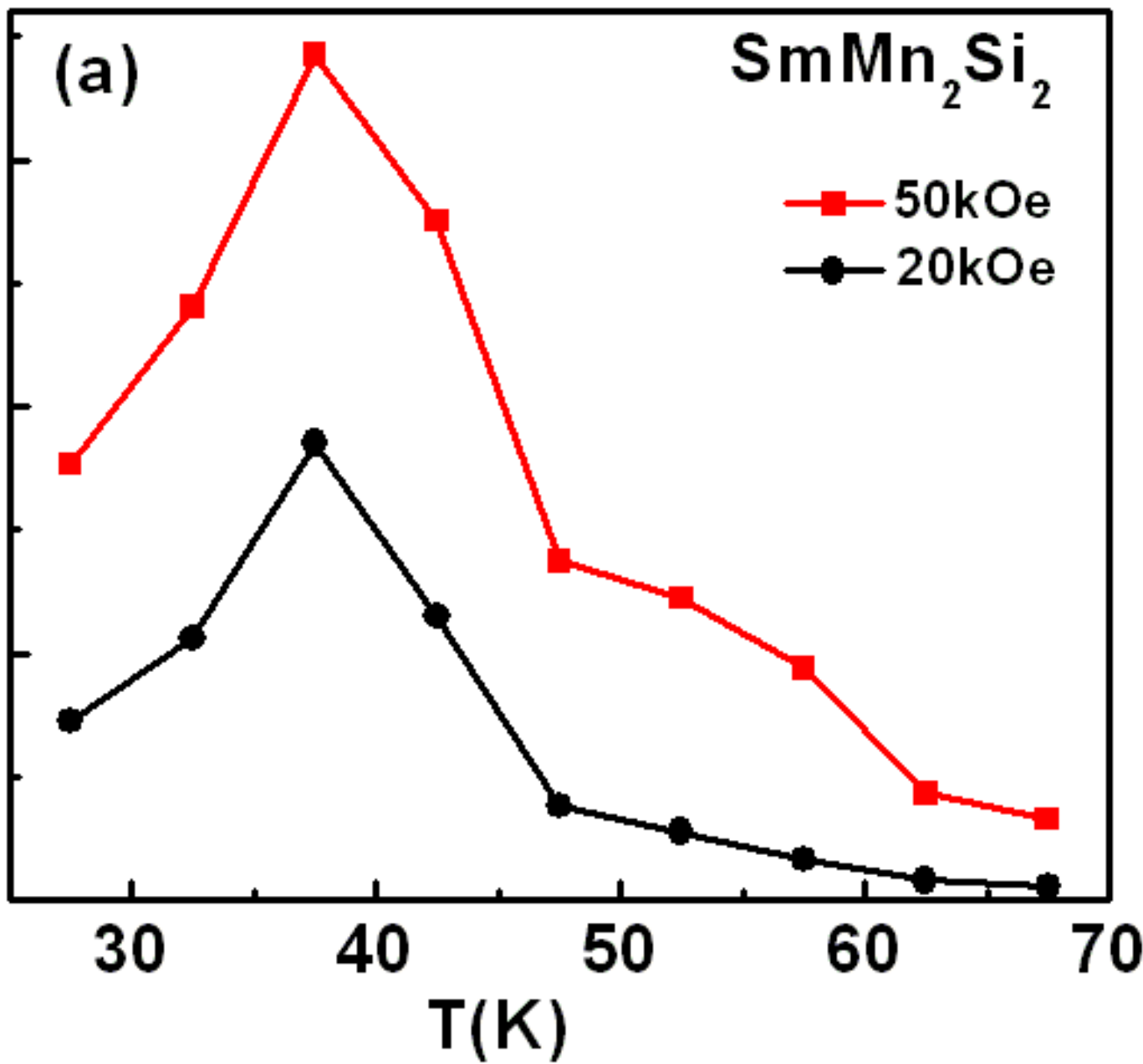

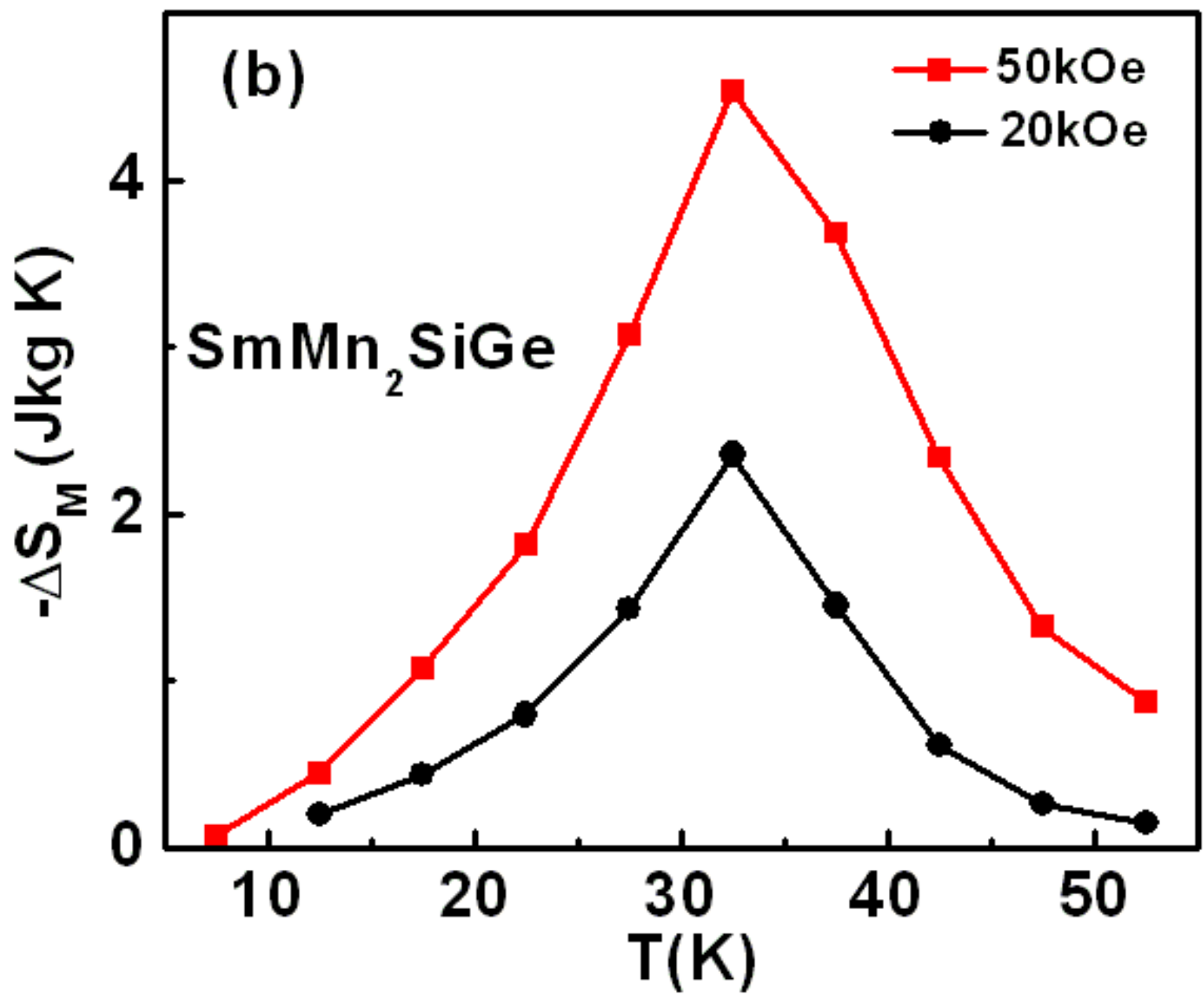

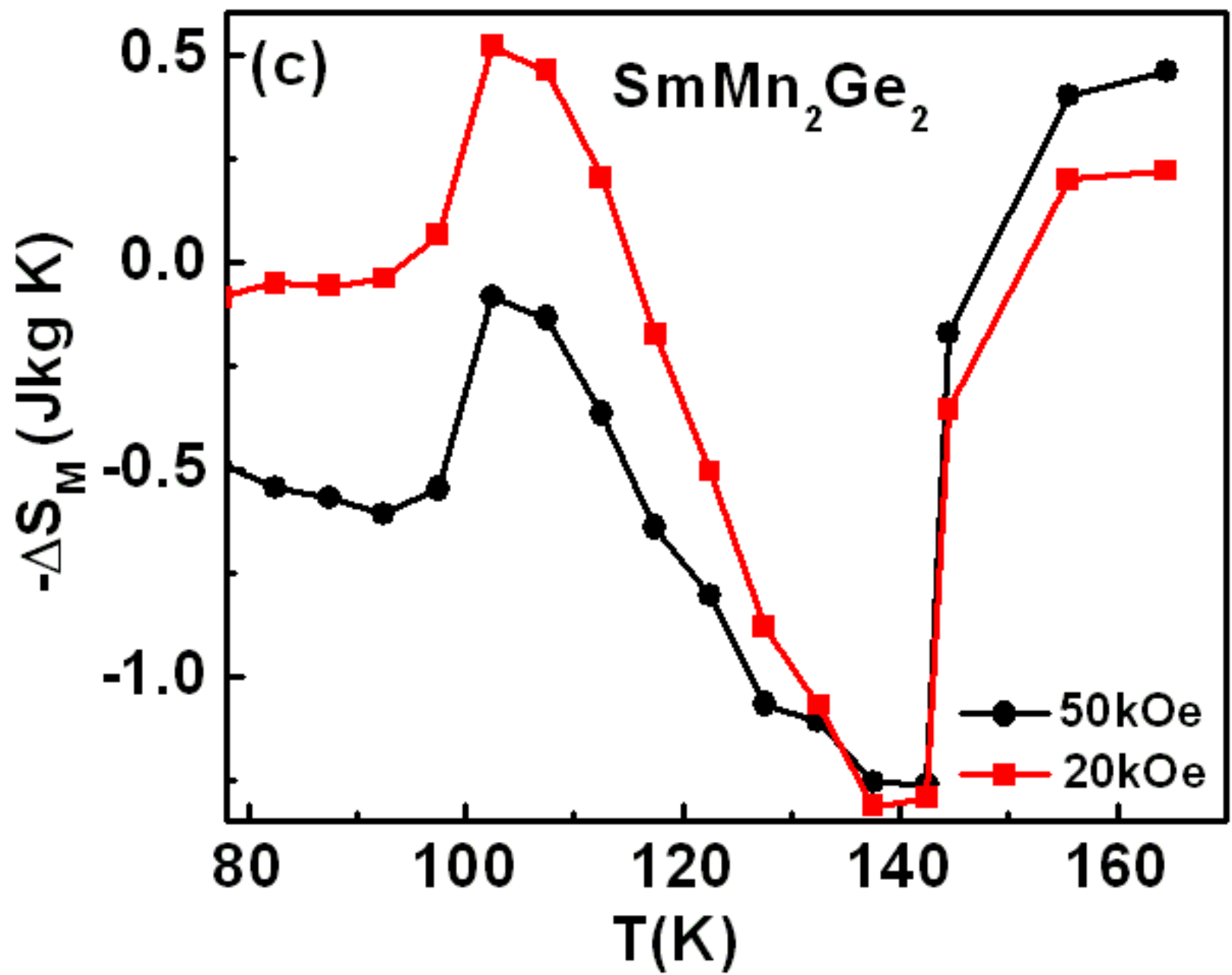

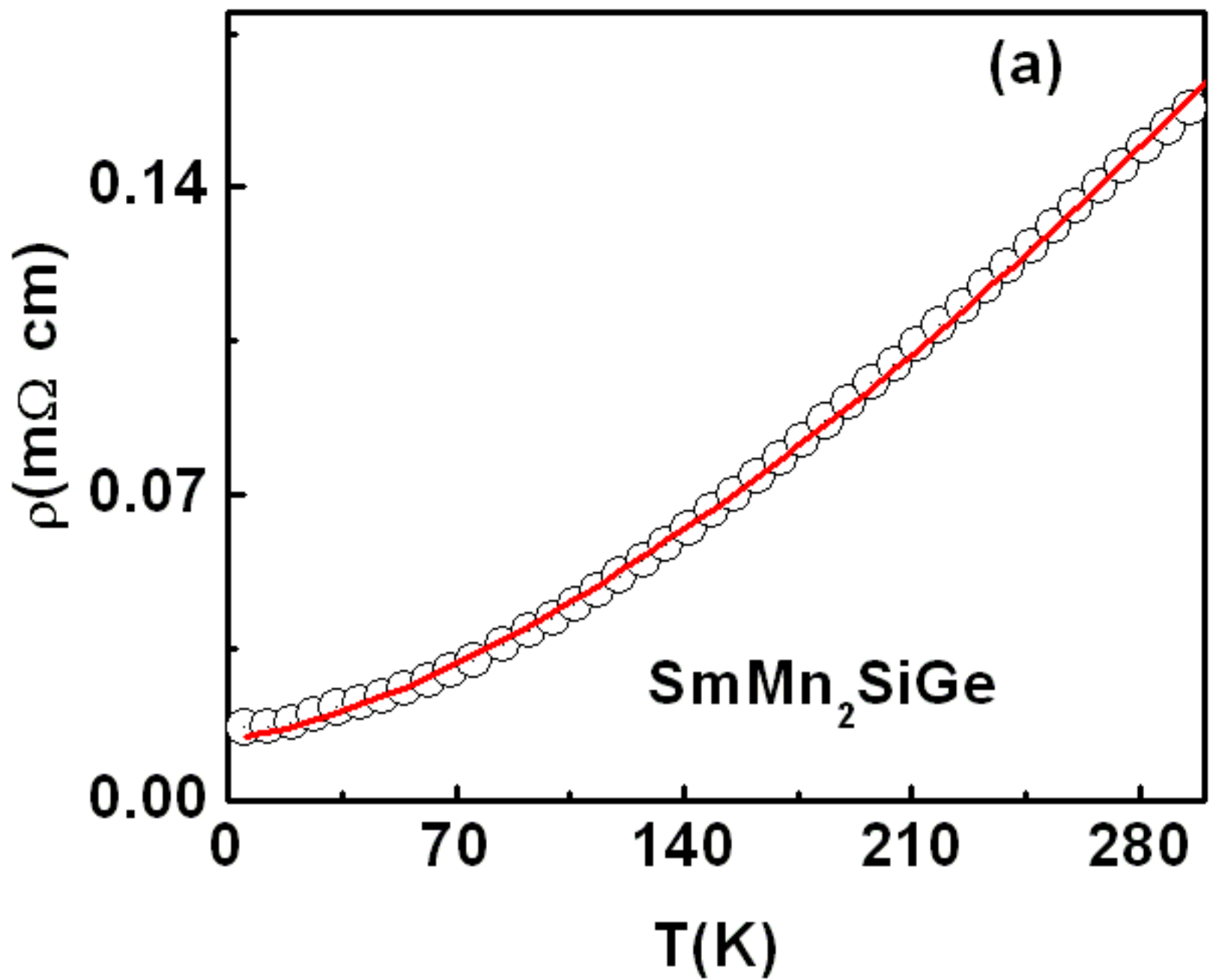

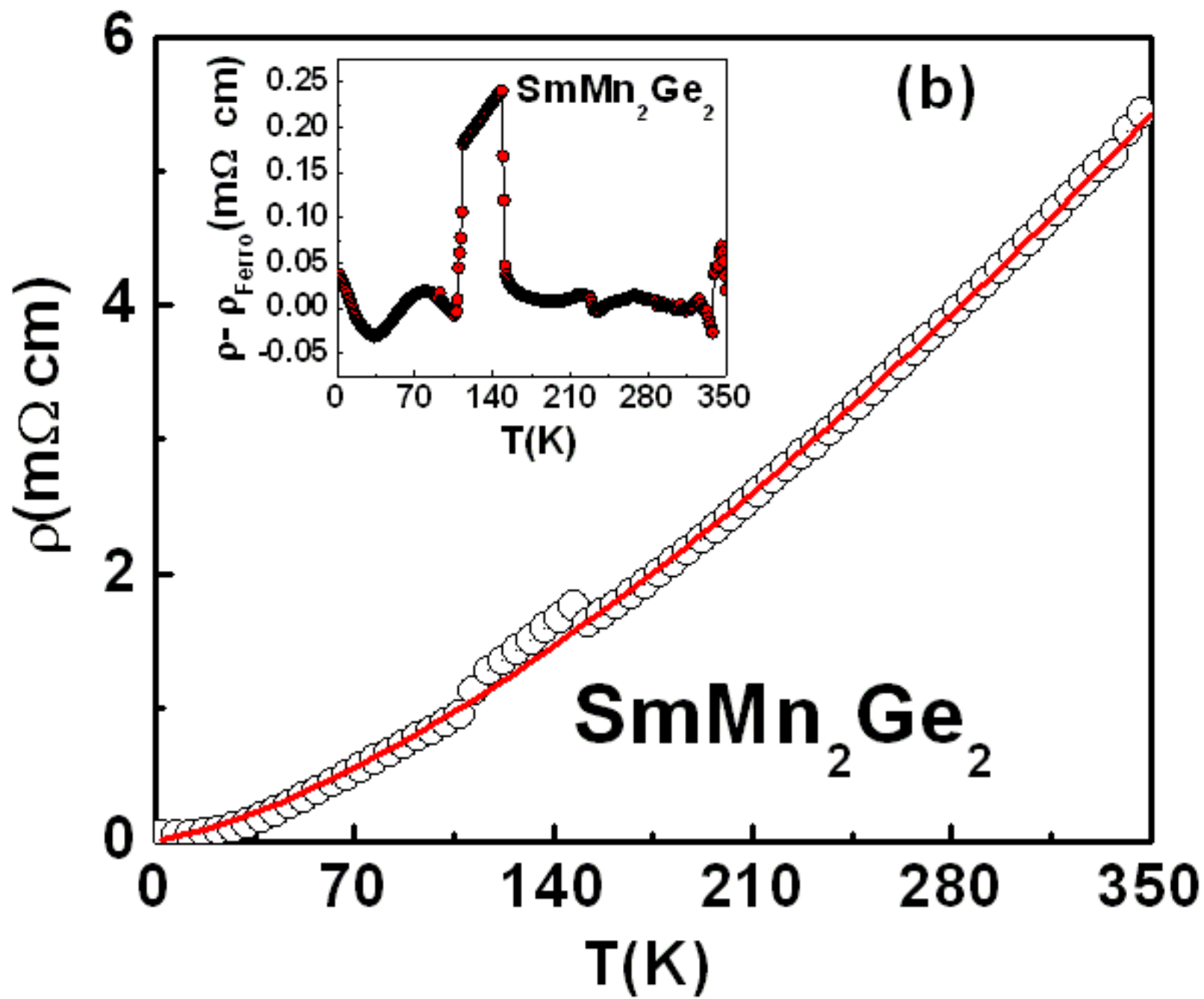